\documentclass[%
 reprint,
 amsmath,amssymb,
 aps,
]{revtex4-2}

\usepackage{bibentry} 
\usepackage{xcolor}
\usepackage{graphicx}
\usepackage{dcolumn}
\usepackage{bm}
\usepackage{amsmath}

\usepackage[mathlines]{lineno}

\begin{document}

\preprint{APS/123-QED}

\title{Natural oscillations of a sessile drop: Inviscid theory}

\author{Saksham Sharma}
 \email{ss2531@cam.ac.uk}
\author{D. Ian Wilson}%
\affiliation{Department of Chemical Engineering and Biotechnology, University of Cambridge,
Philippa Fawcett Drive, Cambridge CB3 0AS, UK
}%





\begin{abstract}
 We present a fully analytical solution for the natural oscillation of an inviscid sessile drop of arbitrary contact angle on a horizontal plate for the case for the case of low Bond number, when surface tension dominates gravity. The governing equations are expressed in terms of the toroidal coordinate system which yields solutions involving hypergeometric functions. Resonant frequencies are identified for zonal, sectoral and tesseral vibration modes. The predictions show good agreement with experimental data reported in the literature, with better agreement than the model of \citeauthor{bostwick} (\textit{J. Fluid Mech.}, vol. 760, 2014, 5-38), particularly for flatter drops (lower contact angle) and higher modes of vibration. The impact of viscous dissipation is discussed briefly.
\end{abstract}

\maketitle


The study of natural oscillations of a drop dates back to \citet{rayleigh}, who presented analytical expressions for the oscillation frequencies of an inviscid, spherical, free drop. \citet{lamb} extended the analysis to include azimuthal mode shapes, using spherical harmonics $Y_l^{m}(\theta , \varphi)$ of degree $l$ and order $m$. \citet{chandra} subsequently considered the contribution of viscosity to explain the damping of the modes of an oscillating viscous drop.

Recent work in this area has shifted from that of a free drop, levitating in air, to a sessile drop on a vibrating flat substrate. While the free drop is generally assumed to be spherical, a sessile drop takes the form of a spherical cap  (when surface tension dominates gravity). The analytical models in the literature either converted the geometry to a simplified form (replacing the planar substrate by a spherical one, \citet{strani}) or developed a solution using spherical coordinates (\citet{bostwick}). While the former approach 
simplifies the models, the latter requires hybrid analytical-numerical schemes. Our interest in the problem lies in developing portable methods for measuring the surface tension of liquids which are only available in small volumes. Measuring the vibration modes of a single droplet allows the surface tension to be estimated. Simplified models introduce uncertainty in such estimates, while hybrid numerical schemes are less attractive for the inverse problem of extracting fluid parameters from modal frequencies: an analytical solution would be preferable. 

\begin{figure}
  \includegraphics[width=0.43\textwidth]{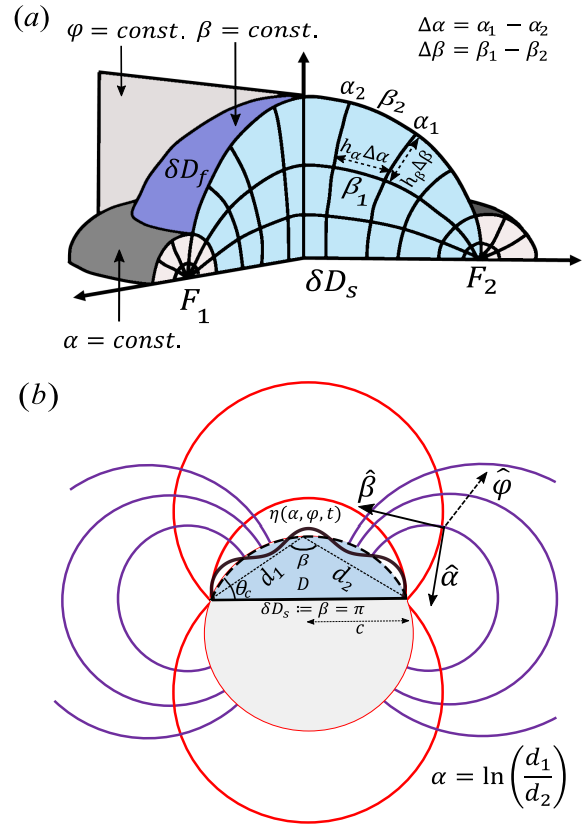}

  \caption{(a) 3D schematic of toroidal coordinate system $\textbf{r}= (\alpha, \beta, \varphi)$ overlaid on a sessile drop. Based on \cite{schematic}. (b) Diametral section of the drop $D$ (blue shaded region) with toroidal gridlines embedded into it. On a red circle, $\beta$ is constant, and on a blue circle, $\alpha$ is constant. Defining expressions for $\alpha$ and $\beta$ are also displayed. The dotted line shows undisturbed interface $\Gamma$ and the solid wavy line represents the perturbation $\eta(\alpha, \varphi,t)$.}
  \label{fig:fig1}
\end{figure}

We present here an analytical solution to this long-standing problem based on the use of toroidal coordinates. The fluid-vapour and fluid-solid boundaries of a spherical cap, $\delta D_{f}$ and $\delta D_{s}$ (cf. Fig.~\ref{fig:fig1}(a)), correspond to a pair of $\beta$-coordinate curves in this system, where the boundary conditions can be directly expressed, without any geometric conversions or complex computations. Solving hydrodynamics equations in this way requires the use of \textit{hypergeometric functions}, which ultimately yields a fully analytical solution in form of Eq. \eqref{eq:final}. \citet{popov} discussed the importance of choosing this framework to solve the sessile drop evaporation problem and we believe this is the first time it has been extended to the vibrating sessile drop. 

\citet{bostwick} (hereafter referred to as Bo-St) presented a hybrid analytical-numerical model which solves the same problem and employs inverse operators to find the solution. Theirs is the most comprehensive investigation of the sessile drop oscillation problem to date. Different types of vibration mode shapes, namely zonal, sectoral and tesseral, were considered which were subsequently validated  experimentally by \citet{bs_2}. The resonant frequencies for the mode shapes discussed by Bo-St are calculated and are compared to the experimental data of \citet{bs_2}. 

The purpose of this Letter is to show that using the toroidal coordinates yields fully analytical solutions for the case of the low Bond number, inviscid sessile drop. We state the hydrodynamic equations with boundary conditions, and perform an eigenmode analysis to find the solution. This model is then used to identify resonant frequencies for zonal, sectoral, and tesseral vibration modes. Its predictions are compared with experimental data reported in the literature. Our model is compared with the Bo-St model and then the possible application of this model to other, related problems is highlighted.

\textit{Theory.\textemdash} The fluid-vapour interface of a sessile drop with contact angle $\theta_{c} \in (0,\pi)$ can be expressed in toroidal coordinates as  $\textbf{r}= (\alpha, \beta, \varphi)$ (cf. Fig.~\ref{fig:fig1}(a)). Variable $\alpha \in [0,\infty]$ varies along the surface $\partial D^{f}$, $\beta \in [0,\pi]$ is the angle subtended by foci $F_{1}, F_{2}$ on $\partial D^{f}$ and $\varphi \in [0,2\pi]$ varies in the azimuthal direction. A small perturbation $\eta (\alpha, \varphi, t)$ on the undisturbed surface $\Gamma$ (with the contact line being fixed) leads to a competition between drop's inertia and capillarity, and the resulting motion is oscillatory in nature (cf. Fig.~\ref{fig:fig1}b). These disturbances, when written in the format of a differential equation, are expressed in terms of 
\begin{equation} \label{eq:scale}
\begin{split}
    h_{\alpha} & =h_{\beta}=\frac{c}{\cosh{\alpha}  - \cos{\beta}}, \\ h_{\varphi} &=\frac{c \, \sin\varphi}{\cosh \alpha - \cos\beta}
\end{split}
\end{equation}
where $c$ is the drop contact radius and $h_{\alpha}$,$h_{\beta}$,$h_{\varphi}$ are the scale factors in the toroidal system.The scale factor gives a measure of change in position of a point on changing one of its coordinates, so a $\Delta \alpha$ change in $\alpha$ (keeping other coordinates constant) corresponds to $h_{\alpha}\Delta \alpha$ change in distance along {$\hat{\alpha}$} (cf. Fig.~\ref{fig:fig1}(a)). 

\textit{Equations and boundary conditions. \textemdash}The flow is assumed to be incompressible and irrotational. The velocity potential $\psi$  satisfies Laplace's equation
\begin{equation} \label{eq:laplace}
    \nabla^{2} \psi = 0 \quad [D]
\end{equation}
in the drop domain D. The equation becomes closed form when subject to the no-penetration condition 
\begin{equation} \label{eq:nopenet}
    \nabla \psi \cdot \bm{\beta} = \frac{1}{h_{\beta}}\frac{\partial \psi}{\partial \beta} = 0   \quad [\partial D_{s}:=  \beta = \pi]
\end{equation}
at the substrate $\partial D^{s}$ and a free-surface kinematic boundary condition 
\begin{equation} \label{eq:kinematic}
    \nabla \psi \cdot \bm{\beta} = \frac{1}{h_{\beta}}\frac{\partial \psi}{\partial \beta}= \frac{\partial \eta}{\partial t} \quad [\partial D^{f} := \beta = \pi - \theta_{c}]
\end{equation}
at the interface $\partial D^{f}$, where the normal velocity is equalized to the time-derivative of perturbation. For an inviscid fluid, applying linear wave theory \cite{lighthill}, the pressure field is described by the momentum equation 
\begin{equation} \label{eq:mom}
    p = - \rho \frac{\partial \psi}{\partial t} \quad[D]
\end{equation}
where $\rho$ is the fluid density. 
A small disturbance $\eta$ to the equilibrium surface $\Gamma$ causes a deviation from the initially spherical shape which is described by the modified Laplace equation \cite[see][p.~105]{myshkis}.
\begin{equation} \label{eq:modlaplace}
    \frac{p}{\gamma} = - (k_{1}^{2}+k_{2}^{2})\eta - \Delta_{T} \eta 
\end{equation}
where $k_{1}$, $k_{2}$ are the principal curvatures, and $\Delta_{T}$ is the Laplace-Beltrami operator.

The first and second fundamental forms of a surface allow the calculation of curvature and Laplace-Beltrami operators respectively, for a parametric surface $\textbf{x}(u^{1}, u^{2})$. The coefficients for first fundamental form are given by the metric tensor.
\begin{equation} \label{eq:metric}
    g_{ij} \equiv \textbf{x}_{i} \cdot \textbf{x}_{j} = \begin{pmatrix} E & F \\ F & G \end{pmatrix}
\end{equation}
\\
where $\textbf{x}_{i}=\frac{\partial \textbf{x}}{\partial u^{i}}$, $\textbf{x}_{j}=\frac{\partial \textbf{x}}{\partial u^{j}}$, $E= \textbf{x}_{i} \cdot \textbf{x}_{i}$, $F= \textbf{x}_{i} \cdot \textbf{x}_{j}$, $G= \textbf{x}_{j} \cdot \textbf{x}_{j}$ and $i,j$ = $1, 2$ (Kreyszig 1991). The derivation of principal curvatures and Laplace-Beltrami operator from the coefficients ($E,F,G$) is given in Appendix A. 

\begin{figure*}
\includegraphics[width=1.0\textwidth]{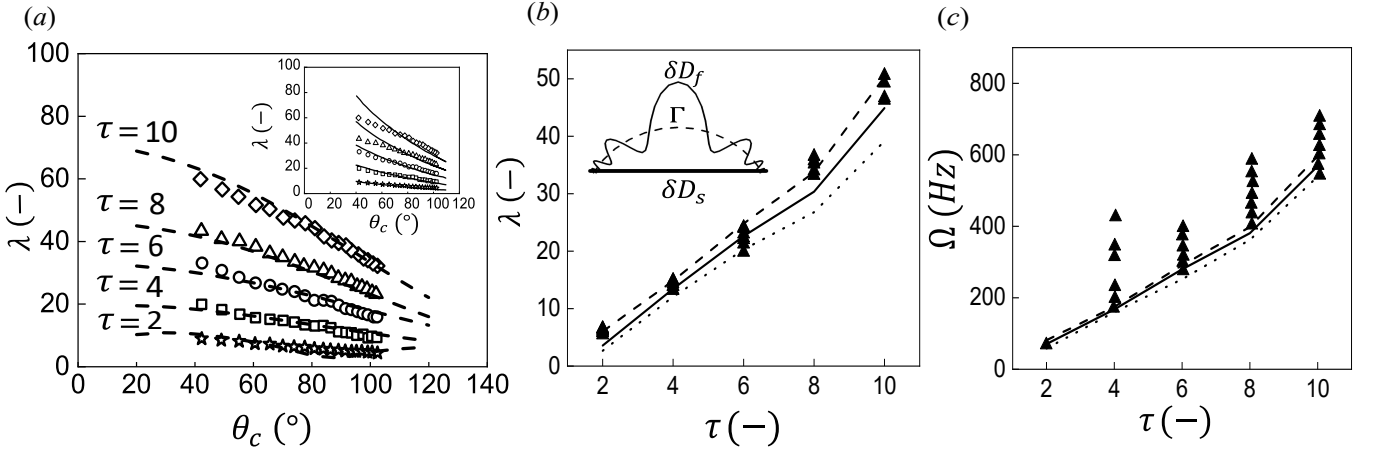}

 \caption{Results for zonal modes, $m=0$. (a) Effect of contact angle $\theta_{c}$ on dimensionless frequency $\lambda$ for toroidal mode number $\tau$. Dotted lines are solutions to Eq. \eqref{eq:final} and symbols are the experimental values reported by \citet{bs_2}. Inset shows the comparison of same experiments with Bo-St model (solid line). (b and c)
 Effect of $\tau$ on $\lambda$ and $\Omega$. Symbols show experimental values reported by (b) \citet{chang_2013} and (c) \citet{chou_2012} for a water drop. Loci show model predictions for different contact angles: (b) solid line, $\theta_{c}=68.6^{\circ}$; dashes, $\theta_{c}=63.6^{\circ}$; dots, $\theta_{c}=73.6^{\circ}$. (c) solid line, $\theta_{c}=79.5^{\circ}$; dashes, $\theta_{c}=68^{\circ}$; dots, $\theta_{c}=91^{\circ}$. Inset in (b) shows interface shape $y$  plotted on $\Gamma$ using Eq. \eqref{eq:final} for $\tau=10$ (not to scale).}
  \label{fig:fig2}
\end{figure*}

\textit{Eigenmode analysis.\textemdash} A drop with pinned contact line is subjected to a small perturbation $\eta(\alpha,\varphi,t)$. Resolving $\eta$ and $\psi$ into individual components: eigenfunctions $y(\alpha)$ and $\phi(\beta,\alpha)$, normal modes (frequency $\Omega$), and azimuthal direction (wavenumber $l$) gives

\begin{equation}\label{eq:eigen}
\begin{split}
        \eta(\alpha, \varphi, t)&=  y(\alpha) e^{i\Omega t} e^{il \varphi} , \\  \psi(\textbf{r}, t)&= \phi(\beta, \alpha) e^{i\Omega t} e^{il \varphi}
\end{split}
\end{equation}

Substituting Eq. \eqref{eq:eigen} into equations Eq. \eqref{eq:laplace}-\eqref{eq:modlaplace} yields
\begin{subequations}\label{eq:tor_eqn}
\begin{alignat}{4}
\frac{\partial}{\partial \beta} \left( h_{\varphi} \frac{\partial \phi}{\partial \beta}\right) + \frac{\partial}{\partial \alpha} &\left( h_{\varphi} \frac{\partial \phi}{\partial \alpha}\right)-\frac{l^{2}}{b \, \sinh \alpha}=0 \quad \label{eq:tor_subeq1} \\          
\frac{\partial \phi}{\partial \beta} &= 0 \label{eq:tor_subeq2} \\ 
\frac{\partial \phi}{\partial \beta} &= \frac{i \lambda y}{b^{2}} \quad \label{eq:tor_subeq3} \\ i \lambda \phi  = 2 \sin^{2}\beta \left(\frac{y}{b}\right) &+ b^{2}\Bigg[\frac{1}{\sinh\alpha}\frac{\partial}{\partial \alpha}\Big(\sinh\alpha \nonumber \\& \frac{\partial}{\partial \alpha}\left(\frac{y}{b}\right)\Big) - \frac{l^{2}}{\sin^{2}(h\alpha)} \frac{y}{b}\Bigg]  \label{eq:tor_subeq4} \\ \lambda^{2}&= \frac{\rho \Omega^{2}c^{3}}{\gamma}\label{eq:tor_subeq5}  
\end{alignat}
\label{straincomponent}
\end{subequations}

where Eq. \eqref{eq:tor_subeq1} is Laplace's equation in toroidal coordinates, Eq. \eqref{eq:tor_subeq2} is the non-penetration boundary condition, Eq. \eqref{eq:tor_subeq3}-\eqref{eq:tor_subeq4} are free surface kinematic boundary conditions and Eq. \eqref{eq:tor_subeq5} gives the scaled frequency $\lambda$.

\textit{Finding the solution.\textemdash} The solution to Laplace's equation Eq. \eqref{eq:tor_subeq1} in toroidal system is given by \cite{lebedev}
\begin{equation}
    \begin{split}
      \phi = & \ a \Big[A P_{v-\frac{1}{2}}^{m}(\cosh\alpha)+ B Q_{v-\frac{1}{2}}^{m} (\cosh\alpha)\Big] \\ & \times [ C \cos{v\beta} +D \sin{v\beta} ]  
    \end{split}
\end{equation}
where $a=\sqrt{2(\cosh \alpha-\cos \beta)}$, $P_{v-\frac{1}{2}}^{m}(\cosh\alpha)$ and $Q_{v-\frac{1}{2}}^{m}(\cosh\alpha)$ are toroidal functions with $v$ as the toroidal degree, $m$ as the azimuthal order. $P_{v}^{m}(z)$ and $Q_{v}^{m}(z)$ (where $z=\cosh\alpha$) are Legendre functions of the first and second kind. At $\alpha=0$, $P_{v}^{m}(1)=1$ and $\lim_{z\to 1+} Q_{v}^{m}(z)=\infty$, which means that the latter is not defined at the apex of the drop. Thus, setting $B=0$ and $v=i\tau$ \cite[see][p.~227]{lebedev} gives
\begin{equation} \label{eq:tor_phieqn}
    \phi = a P_{i\tau-\frac{1}{2}}^{m}(\cosh\alpha)  [C \cos i\tau\beta + D \sin i\tau\beta]
\end{equation}

Using Eq. \eqref{eq:tor_eqn} and substituting $\beta=\pi$ in  Eq. \eqref{eq:tor_subeq2} gives $C = - iD \coth\pi$. Constant $D$ cancels out later in LHS and RHS of Eq. \eqref{eq:final}, hence $D$ and scripts from $P_{i\tau-\frac{1}{2}}^{m}(\cosh\alpha)$ are dropped to re-write Eq. \eqref{eq:tor_phieqn} in a simpler form, as 
    \begin{equation}\label{eq:tor_smpleqn}
        \phi = a P(\cosh\alpha) f(\tau\beta) 
    \end{equation}
where $f(\tau\beta)=\sinh\tau\beta-\coth\tau\pi \cosh\tau\beta$.    
Substituting Eq. \eqref{eq:tor_smpleqn} in Eq. \eqref{eq:tor_subeq3} gives
\begin{equation} \label{eq:disturbance}
    i\lambda \frac{y}{b}= b \frac{\partial \phi}{\partial \beta}=  P  \, b \, (2 \sin\beta f a' + \tau a f') = P T 
\end{equation}
where $a'=1/(2a)$ and $T(\alpha,\beta)= b(2 \sin\beta f a' + \tau a f')$. Functions $T(\alpha,\beta)%
$, $f(\tau \beta)$, $f'(\tau \beta)$ and $P(\cosh\alpha)$ are written without arguments for clarity. 
Substituting the above equation in Eq. \eqref{eq:tor_subeq4} gives, at $\beta=\beta_{0}$; 
\begin{equation}
    \begin{split}
    -\lambda^{2}\phi = \ &2\sin^{2}\beta_{0}PT +  b^{2}\Bigg[\frac{1}{\sinh\alpha}\frac{\partial}{\partial \alpha}\\& \left(\sinh\alpha \frac{\partial}{\partial \alpha}\left(PT\right) \right)   - \frac{l^{2}}{\sin^{2}(h\alpha)} PT \Bigg]
    \end{split}
\end{equation}
This can be re-arranged to
\begin{equation} \label{eq:prefinal}
    -\lambda^{2} \phi = 2 \sin^{2}\beta_{0} PT + b^{2}[I T + II]
\end{equation}
where I and II are  
\begin{subequations}
\begin{gather}
    I = \frac{1}{\sinh\alpha}\frac{\partial}{\partial \alpha} \left(\sinh\alpha \frac{\partial P}{\partial \alpha}  \right) - \frac{l^{2}}{\sin^{2}h\alpha} P \label{eq:symm}\\
    II = \frac{\partial T}{\partial \alpha}\frac{\partial P}{\partial \alpha}+ \frac{1}{\sinh\alpha}\frac{\partial}{\partial\alpha}\left(\sinh\alpha \, P \,  \frac{\partial T}{\partial \alpha} \right)
\end{gather}
\end{subequations}
The term $I$ is equivalent to $(v^{2}-\frac{1}{4})P$ \cite[see][p.~224]{lebedev}. An analogous simplification is performed while deriving an expression for the eigenfrequencies of a free spherical drop in Rayleigh's derivation, \cite[see][p.~246]{landau}. 
Further simplification of the RHS of Eq. \eqref{eq:prefinal} gives 
\begin{equation}\label{eq:final}
    \begin{split}
        -\lambda^{2}= &\left[2sin^{2}\beta_{0}-b^{2}\left( \tau^{2}+\frac{1}{4} \right) \right]\frac{T}{af} \\ & + \frac{b^{2}}{af}\left[T'\left(\frac{P'}{P}(1+\sinh\alpha)+\coth\alpha\right)+T'' \right]
    \end{split}
\end{equation}
where the derivatives are w.r.t. $\alpha$. The expressions for $T', T'', P, P'$ (which falls under the class of \textit{hypergeometric functions}) are given in Appendix B. 

\begin{figure*}
 \includegraphics[width=1.0\linewidth]{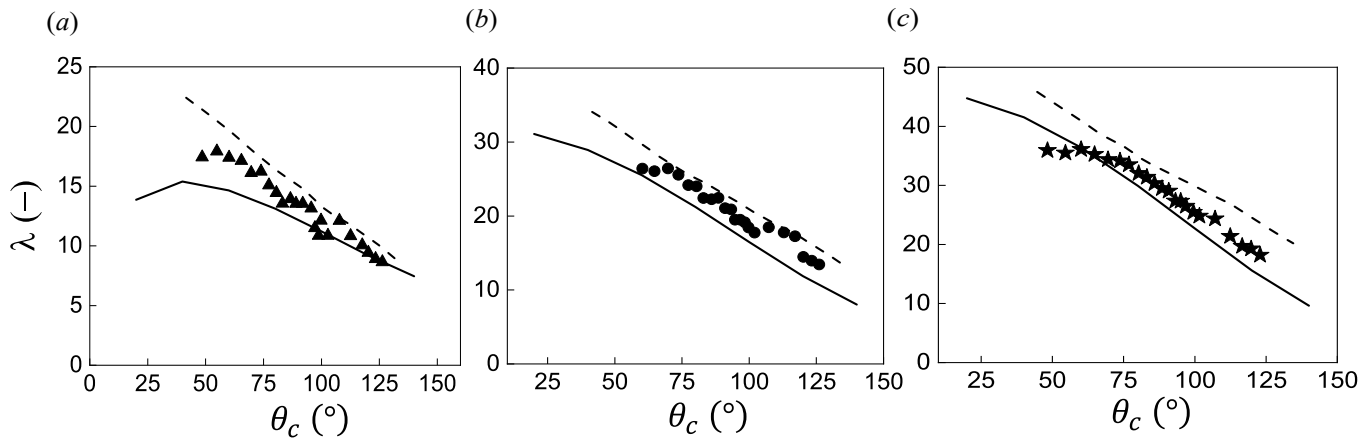}
\caption{Effect of contact angle on dimensionless frequency for sectoral modes, $\tau=m$, for (a) [5,5], (b) [7,7], (c) [9,9]. Solid loci show the solutions to Eq. \eqref{eq:final}), dashed loci are the results presented by \citet{bostwick}. Symbols indicate experimental data reported by \citet{bs_2}.}
\label{fig:fig4}
\end{figure*}

\textit{Results.\textemdash} The variation of dimensionless frequency $\lambda$ with contact angle  $\theta_c = \pi-\beta_0$ is determined by solving Eq. \eqref{eq:final}. Previous studies such as Bo-St classified the vibrational modes as zonal ($m=0$), sectoral ($\tau=m$) and tesseral ($m\neq 0, \tau\neq m$). Results are presented for each type of mode in turn.

When the disturbance of the interface is axisymmetric, the mode shapes are termed zonal. For a sessile drop of fixed contact radius $c$, increasing the contact angle $\theta_{c}$ increases the volume of drop (inertia) and thus decreases the frequency $\lambda$ (cf. Fig. \ref{fig:fig2}(a)). There is good agreement between the model and the data of \citet{bs_2}, particularly at higher mode numbers. For instance, for $\tau=10$ and $\theta _{c}=40^{\circ}$, our model overpredicts slightly by a factor of 1.05, while the Bo-St model overpredicts by a factor of 1.25 (c.f. inset Fig.\ref{fig:fig2}(a)). For the other modes at $\theta _{c}<60^{\circ}$, agreement is quite better than Bo-St model. For the lowest mode ($\tau=2$), our model underpredicts  $\lambda$ slightly for $\theta_{c}$ around 90$^{\circ}$. This arises from the calculation, where there is a transition from real to imaginary values at 90$^{\circ}$.  Higher mode numbers correspond to more points (nodes) of intersection of the disturbed interface $\delta D_{f}$ with the undisturbed interface $\Gamma$. Since there is no variation in the azimuthal direction, a front view (cf. inset Fig. \ref{fig:fig2}(b)) is sufficient to describe the mode shape. This Figure shows the case of 10 nodes ($\tau=10$).

Fig. \ref{fig:fig2}(b,c) shows further comparisons of zonal modes with data sets reported by \citet{chou_2012} and \citet{chang_2013}. In Fig. \ref{fig:fig2}(b), the experimental values fall within the range of theoretical frequencies calculated for the range of contact angles $\theta_{c}$ involved. Here for the higher modes, $\tau$ = 8 and 10, the frequencies fall on upper part of theoretical span because of limited number of data points, available only for droplets $\gtrsim
5\mu L$ \citep[see][Fig.4(a)]{chou_2012}, whereas lower modes are experimentally discernible even for smaller droplets. Also, there is a slight increase in slope at $\tau=8$ which is also apparent in Fig. \ref{fig:fig2}(b), in the form of slight over-prediction of the model, at around $\theta _{c}=60^{\circ}$. In Fig. \ref{fig:fig2}(c), the width of the predicted frequency band is small and lies at the lower end of the spectrum of observed frequencies. One explanation for this could be because the viscous effects are neglected. \citet{chang_2013}  reported that the bandwidth of predicted frequencies increased when viscous contributions were added (noting that the dimensional frequency is plotted here). \citet{bs_2} showed that the viscous contribution is characterised by the Ohnesorge number, $Oh=\mu/\sqrt{\rho c \gamma}$, and even a small value of $Oh=0.003$ for water at $25^{\circ}$C (instead of $Oh=0$ for the inviscid case) altered the resonance peak from an infinite to a finite value and thus increased the bandwidth of predicted frequency (see \citet[p.~446]{bs_2}). 
\begin{figure*}
  \includegraphics[width=1.0\linewidth]{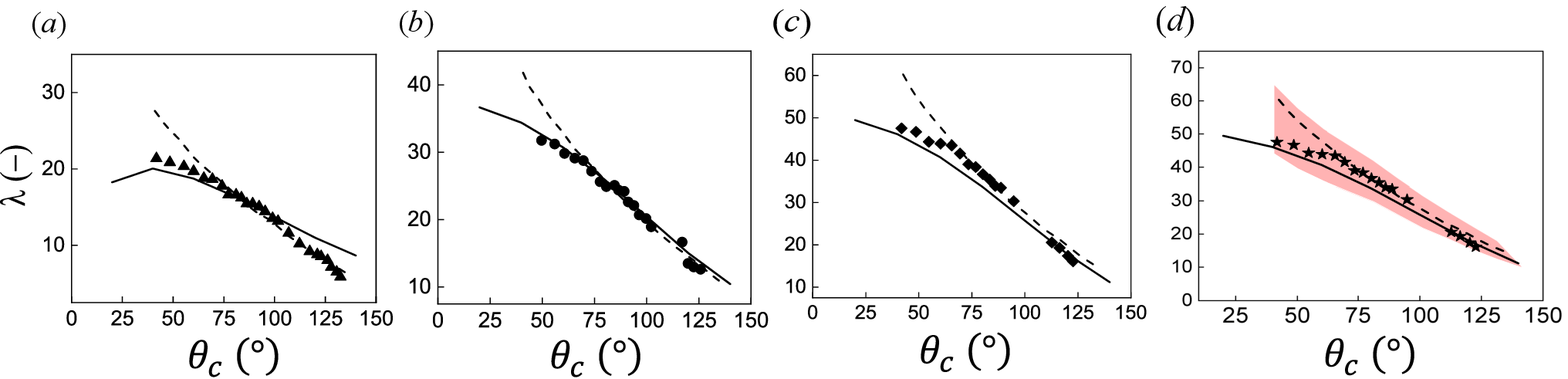}
  \caption{Effect contact angle on dimensionless frequencies for tesseral modes with $[\tau,m]$ values of (a) [5,3], (b) [7,5], (c) [9,5] and (d) [9,7]. Solid loci show the solutions to Eq. \eqref{eq:final}), dashed lines are the results presented by \citet{bostwick}. Symbols show experimental data reported by \citet{bs_2}. The shaded region in panel (d) represents the range of frequencies calculated using VPF theory by \citet{bs_2} for water, with substrate forcing and viscosity included.} 
  \label{fig:fig5}
\end{figure*}                                           A non-axisymmetric mode with wavenumber pair $[\tau,m]$ has $m$ longitudinal intersections and $(\tau-m)/2$ latitudinal intersections (or $\tau-m$ nodes on the interface) with the undisturbed interface $\Gamma$ \cite[p.~19]{bostwick}. A sectoral mode, with $\tau=m$, is a special case where there are only longitudinal intersections. Fig. \ref{fig:fig4} compares the experimental frequencies reported by \citet{bs_2} with our model and the Bo-St model. The latter tends to overpredict $\lambda$ whereas Eq. \eqref{eq:final} tends to underpredict the experimental values. There is fairly good agreement with our model for $\tau=9$. For $\tau=$ 5 and 7, the two models bracket the data. Our model predicts a local maxima at $45^{\circ}$ for $\tau=5$; this remains to be verified with experimental data over a wider range of $\theta_{c}$ values. \par  A tesseral mode shape with wavenumber pair $[\tau,m]$ has non-zero longitudinal and latitudinal intersections because $\tau \neq m$. Fig.~\ref{fig:fig5} compares the results for our model and the Bo-St model in a similar fashion to the sectoral mode. For the $\tau=9$ cases, our model shows a slight underprediction and agrees with the experimental data fairly well for all $\theta_{c}$ values investigated: at contact angles $\leq65^{\circ}$ the Bo-St model does not capture the observed trend and overpredicts $\lambda$. The superior peformance of our model for flatter drops (lower $\theta_{c}$) is attributed to the use of toroidal coordinates, which fit the sessile drop naturally. For $\tau=7$ there is good agreement with both models until smaller $\theta_{c}$ for Bo-St. Neither model captures the observed behaviour for $\tau=5$: our model captures the frequencies at low $\theta_{c}$ while the \citet{bostwick} model is superior at higher values in this case. \par \textit{Discussion.\textemdash} There is generally good agreement between the solutions calculated using the toroidal coordinate framework and experimental data sets reported in the literature. There are exceptions, e.g. Fig. 4(a), 5(a), and we here consider whether the mismatch between the predictions of the model and the experiments could arise from the assumptions made in obtaining Eq. \eqref{eq:final}. \par The model considers the sessile drop on a substrate as a mass-spring system. Viscous effects and substrate-drop interactions are neglected. These assumptions were also made in the Bo-St model and were subsequently relaxed in the work of \citet{bs_2}. Addition of viscous effects changed the system to a mass-spring-damper system while the substrate-drop interactions were modeled as Faraday oscillations by assuming the forcing to be occurring through the bulk pressure in the drop. \citet{bs_2} incorporated these effects using VPF (viscous potential flow) theory, increasing the computational expense of the calculations. The envelope of solutions calculated using VPF is shown in Fig.~\ref{fig:fig5}(d) alongwith the solutions from inviscid Bo-St model and the current work for $[\tau,m]$ = [9,7]. The envelope spans both inviscid models and again does not capture the behaviour at low $\theta_{c}$ well.  It is expected that the addition of viscous and substrate contributions to the model here will modify Eq. \eqref{eq:final} and increase the bandwidth of predicted frequencies. This is the subject of ongoing work, where the aim is to identify the contributions of viscous damping, contact angle mobility (from pinned to mobile) and substrate forcing, and thereby establish when significant differences will arise from the inviscid model. \par                    Which interesting problems can this model tackle? A possible application could be in understanding the vibration induced ejection of a lone droplet from a sessile drop. Experiments by \cite{vukasinovic} demonstrated that the dimensionless acceleration threshold $a_{c}$ of the substrate when drop is ejected scaled with the dimensionless forcing frequency $\omega_{c}$ as $a_{c} \sim \omega_{c}^{1.04}$, for a drop of volume $5\mu L$ (low-mode excitation). The exponent increased to 1.21 on increasing the volume to $15\mu L$ and approached the limit $4/3$ for large volumes. This higher limit is characterised as high-mode excitation where drop-substrate coupling is less pronounced, hence a simple scaling theory, such as \cite{goodridge}, explains it easily. However, in the case of low-mode excitation, substrate-drop coupling causes a direct influence of drop volume on the scaling relation. In this case, our toroidal framework (with viscous and substrate contributions) can be helpful to derive a scaling relation with geometrical factors (arising from the spherical cap) included. It should be noted that the toroidal description is not suitable for cases the effect of gravity is significant. In such cases, a confocal ellisoidal coordinate system can be useful, where the drop interface can be assumed elliptical \cite{lubarda}. \par \textit{Conclusions.\textemdash} A new method for calculating the resonant frequencies of a vibrating sessile drop is presented. We solved the governing hydrodynamics equations (\ref{eq:laplace}-\ref{eq:modlaplace}) using an eigenmode reduction approach within a toroidal coordinate framework. The shift to toroidal coordinates from spherical systems employed previously makes the derivation fully analytical and the result, Eq. (\ref{eq:final}), is readily evaluated. \par          The predicted frequencies for zonal, sectoral and tesseral mode shapes show fairly good agreement with experimental data sets reported in the literature, especially for higher modes. The agreement is better in several cases than the model of \citet{bostwick} used as a benchmark, particularly for small contact angles.  The discrepancies between experiments and predictions could possibly be accounted for by including viscous and substrate contributions, but this is likely to require numerical solution.                              

In a broader context, coordinate transformations from a Cartesian to a toroidal framework Eq. \eqref{eq:transform} constitutes a type of conformal mapping because the angle between the curves or gridlines ($90^{\circ}$) is preserved. This technique is very powerful for solving physical problems with complicated geometries, such as water waves over a variable bottom \cite{fokas} and Hele-Shaw flows \cite{richardson}. 

To summarise, our model provides a concise solution to the sessile drop vibration problem which opens a new window to the researchers interested in this and related problems. Drop resonance is of interest to those who move an inclined drop by vibrating the plane \cite{deegan}, and the splitting of a drop into smaller drops using acoustic fields.

We wish  to thank Dr. R.K. Bhagat, Dr. H. Tankasala, A.J.D. Shaikeea for fruitful discussions on this problem. Funding for S.S. from the Cambridge India Ramanujan Scholarship is gratefully acknowledged. 

\textit{Appendix A (Differential geometry of toroidal system)\textemdash} A general point on the surface $\beta =\beta_{0}$ is $\textbf{x}(\alpha,\varphi)= (x,y,z)$ such that
\begin{equation} \label{eq:transform}
    x = \frac{c \, \sinh\alpha \, \cos\varphi}{b} , y = \frac{c \, \sinh\alpha \, \sin\varphi}{b}, z = \frac{c  \, \sin\beta}{b} 
\end{equation}
where $b = \cosh\alpha - \cos\beta$ \cite{lebedev}. Putting $i=\alpha$, $j=\beta$ in Eq. \eqref{eq:metric} gives

\begin{equation}
\begin{split}
    \textbf{x}_{\alpha} &= \begin{pmatrix} \frac{c \cos\varphi \,d }{b^{2}} , \frac{c \sin\varphi \, d }{b^{2}} , -\frac{c \sin\beta \, \sinh\alpha}{b^{2}} \end{pmatrix} \\
\textbf{x}_{\varphi}&= \begin{pmatrix} \frac{-c \sinh\alpha \, \sin\varphi}{b}  , \frac{c \sinh\alpha \cos \varphi }{b} , 0 \end{pmatrix} \\
E &= \textbf{x}_{\alpha} \cdot \textbf{x}_{\alpha} = \Big(\frac{c}{b}\Big)^{2}\\
F &= \textbf{x}_{\alpha} \cdot \textbf{x}_{\beta} = 0 \\
G &= \textbf{x}_{\beta} \cdot \textbf{x}_{\beta} =\Big(\frac{c\, \sinh\alpha}{b}\Big)^{2}\\
W &= \sqrt{EG-F^{2}}= \frac{c^{2} \, \sinh\alpha}{b^{2}}
\end{split}
\end{equation} 
where $d=b \ (\cosh\alpha-\sin^{2}h\alpha$) and $W$ is the determinant of the metric tensor. \\
The coefficients of second fundamental form of surface are:  $L=\textbf{x}_{ii} \cdot \textbf{n}$, $M=\textbf{x}_{ij} \cdot \textbf{n}$, $N=\textbf{x}_{jj} \cdot \textbf{n}$ where $\textbf{n}=\frac{\textbf{x}_{i} \times \textbf{x}_{j}}{|\textbf{x}_{i} \times \textbf{x}_{j}|}$
and $|\textbf{x}_{i} \times \textbf{x}_{j}| = W$. Again, $i=\alpha$, $j=\beta$ gives

\begin{equation}
\begin{split}
    L &= \frac{ \left(\textbf{x}_{\alpha \alpha} \textbf{x}_{\alpha} \textbf{x}_{\beta}\right)}{W} =  \frac{-c \, \sin \beta }{b^{2}} \\
    M &= \frac{ \left(\textbf{x}_{\alpha \beta} \textbf{x}_{\alpha} \textbf{x}_{\beta}\right)}{W} =0 \\
    N &= \frac{ \left(\textbf{x}_{\beta \beta} \textbf{x}_{\alpha} \textbf{x}_{\beta}\right)}{W} =  \frac{-c \, \sin \beta \, \sinh^{2} \alpha}{b^{2}} 
\end{split}
\end{equation}
In Eq. \eqref{eq:modlaplace}, the first term in RHS 
\begin{equation}
    k^{2}_{1}+k^{2}_{2}= \left(\frac{EN-2FM+GL}{W^{2}} \right)^{2} - 2 \frac{LN-M^{2}}{W^{2}} = \frac{2 \sin^{2}\beta}{c^{2}}
\end{equation}
and second term (Laplace-Beltrami operator) is 
\begin{equation}
\begin{split}
    \Delta_{T}\eta & = \frac{1}{W} \left[ \frac{\partial}{\partial \alpha} \left(\frac{G\eta'_{\alpha}-F\eta'_{\varphi}}{W} \right) + \frac{\partial}{\partial \varphi} \left(\frac{E\eta'_{\alpha}-F\eta'_{\varphi}}{W} \right) \right] \\
    & = \frac{b^{2}}{c^{2}} \Big[ \frac{1}{\sinh\alpha} \frac{\partial}{\partial \alpha} \big( \sinh\alpha \frac{\partial \eta}{\partial \alpha} \big) + \frac{1}{\sin^{2}h\alpha} \frac{\partial}{\partial \varphi}\big(\frac{\partial \eta}{\partial \varphi}\big)\Big]
\end{split}
\end{equation}
where the formula and notations are followed from \citet[p.~105]{myshkis}.

\textit{Appendix B (Hypergeometric functions).\textemdash} Hypergeometric functions are solutions to the second order ODE encountered while using a system of orthogonal curvilinear coordinates to solve Laplace's equation \cite[see][p.~161-173]{lebedev}. In our case, we use toroidal system to solve Laplace's equation and find Legendre functions of the first kind $P_{v}(z)$ as the solution (hence, referred to as the toroidal functions). The integral representations of these functions are given below, for different cases:

\begin{enumerate}
\item When $m=0$ \cite[see][p.~173]{lebedev} \\
\begin{equation}  P_{v}(\cosh \alpha)= A_{1} \int_{0}^{\infty} \frac{\cosh (v + \frac{1}{2})\theta}{\sqrt{2 \cosh \theta + 2 \cosh \alpha}} d\theta 
\end{equation}
for $\alpha > 0, -1< Re(v) < 0$ and 
\begin{equation}
 P_{v}'(\cosh \alpha)= A_{1} \int_{0}^{\infty} \frac{-\cosh (v + \frac{1}{2})\theta}{(2 \cosh \theta + 2 \cosh \alpha)^\frac{3}{2}} d\theta, 
\end{equation}
for $\alpha > 0, -1< Re(v) < 0$. The derivative sign is w.r.t. $\alpha$ and $A_{1} =\frac{2}{\pi} \cos \left(v + \frac{1}{2}\right)\pi $.
\item When $m\neq0$ \cite[see][p.~172,199]{lebedev} 

\begin{equation} \label{eq:Pvm_der}
     P_{v}^{m}(\cosh \alpha)= A_{2} \int_{-\alpha}^{\alpha} \frac{ e^{-(v+\frac{1}{2})\theta} T_{m}(\cos \psi)}{\sqrt{2 \cosh \alpha - 2 \cosh \theta}} d\theta
\end{equation}
where $A_{2}=\frac{\Gamma(v+m+1)}{\pi \Gamma(v+1)}$, $\Gamma$ is the gamma function and $T_{m}$ is the Chebyshev polynomial. 
The derivative of Eq. \eqref{eq:Pvm_der} is evaluated by numerically measuring the slope of curve.

\end{enumerate}

Other functions used in Eq. \eqref{eq:final} are
\begin{equation}
\begin{split}
    \frac{T'}{a f} = \ &b' \left(2 \sin \beta_{0} \frac{a'}{a} + \tau \frac{f'}{f}\right) \\ & +b \left(2 \sinh\alpha  \left(\sin\beta_{0} \frac{a''}{a} + \tau \frac{f'}{f} \frac{a'}{a} \right) \right)
\end{split}
\end{equation}

\begin{equation}
\begin{split}
    \frac{T''}{a f} = \ &2b' \sinh \alpha \left(2 \sin \beta_{0} \frac{a'}{a}+\tau \frac{f'}{f} \right)  \\& + b' \left(2 \sinh\alpha \left(2 \sin\beta_{0} \frac{a''}{a} + \tau \frac{f'}{f} \frac{a'}{a}\right)\right) \\&+  4b \sinh^{2} \alpha \left(2 \sin\beta_{0}\frac{a'''}{a}+\tau \frac{f'}{f}\frac{a''}{a}\right) \\&+ b'' \left(2 \sin \beta_{0} \frac{a'}{a}+\tau \frac{f'}{f} \right) \\&+ 2b \cosh\alpha \left(2 \sin\beta_{0}\frac{a''}{a}+\tau \frac{f'}{f}\frac{a'}{a}\right)
\end{split}
\end{equation}

\bibliography{apssamp}

\end{document}